\documentclass[a4paper,11pt]{article}
\pdfoutput=1
\usepackage{jheppub} % for details on the use of the package, please
\usepackage{axodraw2}

\title{The Case for Future Hadron Colliders From $B \to K^{(*)} \mu^+ \mu^-$ Decays} 
\author[a]{B.C. Allanach,}
\author[b]{Ben Gripaios,}
\author[a,b]{Tevong You\footnote{Corresponding author.}}

\affiliation[a]{DAMTP, University of Cambridge, Wilberforce Road, Cambridge, 
CB3 0WA, United Kingdom}
\affiliation[b]{Cavendish Laboratory, University of Cambridge, J.J. Thomson Avenue, Cambridge, CB3 0HE, United Kingdom}

\emailAdd{B.C.Allanach@damtp.cam.ac.uk}
\emailAdd{gripaios@hep.phy.cam.ac.uk}
\emailAdd{tty20@cam.ac.uk}
\preprint{Cavendish-HEP-2017-11,  DAMTP-2017-39}
\abstract{
Recent measurements in $B \to K^{(*)} \mu^+ \mu^-$ decays are somewhat
discrepant with Standard Model predictions. They may be harbingers of new
physics at an energy scale potentially accessible to direct discovery. We
estimate the sensitivity of future hadron colliders to the possible new
particles that may be responsible for the anomalies at tree-level: leptoquarks
or 
$Z^\prime$s. We consider luminosity upgrades for a 14 TeV LHC, a 33 TeV LHC,
and a 100 TeV $pp$ collider such as the FCC-hh. In the most conservative and pessimistic models, for narrow particles with perturbative couplings, $Z^\prime$ masses up to 20 TeV and leptoquark masses up to 41 TeV may in principle explain the anomalies. Coverage of $Z^\prime$ models
is excellent: a 33 TeV 1 ab$^{-1}$ LHC is expected to cover most of the parameter space up to 8 TeV
in mass, whereas the 100 TeV FCC-hh with 10 ab$^{-1}$ will cover all of it. A
smaller portion of the leptoquark parameter space is covered by future
colliders: for example, in a $\mu^+\mu^-jj$ di-leptoquark search, a 100 TeV
10 ab$^{-1}$ collider has a projected sensitivity up to leptoquark masses of
12 TeV (extendable to 21 TeV with a strong coupling for single leptoquark
production).
}

\begin{document}
\maketitle 
\flushbottom

\section{Introduction}

Perhaps the most convincing anomalies\footnote{Here, `anomaly' refers to a
  measurement that is discrepant with respect to a Standard Model prediction.}
observed in the LHC data thus far are
those seen in ratios of branching ratios of semi-leptonic $B$-to -$K$ or
-$K^*$ decays in LHCb~\cite{Aaij:2014ora, Aaij:2017vbb}.  Though they involve
sensitive measurements of rare processes, they are theoretically
clean~\cite{Hiller:2003js} and apparently a clear signal of violation of
lepton universality, a principle that is sacrosanct in the gauge interactions
of the Standard Model (SM). Moreover, the fact that such processes arise only
at loop level in the SM means that, even though the observed deviations are
large compared to the SM contribution, they could plausibly be explained by
tree-level exchange of new particles at the TeV scale, with couplings of
comparable size to those present in the SM.\footnote{In contrast, apparent
  deviations seen in $B$-to-$D$ decays~\cite{Lees:2012xj, Lees:2013uzd,
    Huschle:2015rga, Sato:2016svk, Hirose:2016wfn, Aaij:2015yra, Aaij:2017uff}
  are comparable in size to SM tree-level contributions, and so seem to call
  for either an implausibly low scale of new physics or rather large
  couplings.} 

To put the measurement of these ratios in context, we summarise some of the
related anomalies that preceded them: the first sign of a discrepancy appeared
in the $P_5^\prime$ observable~\cite{DescotesGenon:2012zf} of angular
distributions in $B \to K^* \mu^+ 
\mu^-$ decays~\cite{Aaij:2014pli, Aaij:2013qta,
  Aaij:2015oid,Descotes-Genon:2013vna,Descotes-Genon:2015uva}, designed in 
such a way that hadronic uncertainties cancel out and are under control. LHCb
found a $3.4 \sigma$ anomaly~\cite{Aaij:2015oid}, supported somewhat at
the $ 2 \sigma$ level by a later BELLE
measurement~\cite{Abdesselam:2016llu}. These were also consistent with a $
3.2 \sigma$ tension in $B_s \to \phi \mu^+ \mu^-$~\cite{Aaij:2015esa}. Indeed,
various global fits including LHCb, Belle, BaBar, CMS, and ATLAS data to a
variety of $b \to s \mu^+\mu^-$ kinematic observables indicated a non-zero
value for a particular Wilson coefficient parameterising new physics coupling
to left-handed quarks and muons, with a statistical pull $\gtrsim 4
\sigma$~\cite{Descotes-Genon:2013wba, Altmannshofer:2013foa, Beaujean:2013soa,
  Hurth:2013ssa, Capdevila:2017bsm, Altmannshofer:2017yso, Hiller:2017bzc,
  Geng:2017svp, Ciuchini:2017mik, DAmico:2017mtc}. However, these observables
could still have been heavily affected by residual theoretical uncertainties
in the SM 
prediction. It was therefore notable that subsequent measurements of the more
theoretically clean  ratios 
$R_K=BR(B \rightarrow K \mu^+ \mu^-)/BR(B \rightarrow K
e^+ e^-)$ and $R_{K^*}=BR(B \rightarrow K^* \mu^+ \mu^-) / BR(B \rightarrow
K^* e^+ e^-)$
both observed deviations at around the 
$2.5 \sigma$ level each~\cite{Aaij:2014ora, Aaij:2017vbb}. Moreover, fits
to these two clean observables alone demonstrate a pull away from the SM at
more than around
$ 4 \sigma$ on the \emph{same} Wilson coefficient as the one from the
global fit to the other observables~\cite{Capdevila:2017bsm,
  Altmannshofer:2017yso, Hiller:2017bzc, Geng:2017svp, Ciuchini:2017mik,
  DAmico:2017mtc, Celis:2017doq}. This non-trivial consistency of the various
anomalies goes some way towards explaining the level of interest in them,
despite the significance of each individual measurement being low.  

Even if the anomalies really are signatures of physics beyond the SM (and
further data or a better understanding of the SM predictions may well indicate
that they are not), we face the problem that the effects we see in $B$ decays
arise indirectly, via exchange of virtual states that are far from being
on-mass-shell. To confirm the presence of new physics, and to begin the long,
but tremendously exciting, programme of exploring Nature's next layer, we will
need to produce the new particles directly on-shell, at a current or future
collider. But in trying to plan for this, we must overcome a serious obstacle:
the size of the effects being seen currently fixes neither the identity, nor
the mass, nor the couplings of the new particles. So, at least without further
consideration, not only do we not know what energy threshold a collider would
need to reach to produce the new states, but also, even if we did know what
energy were needed, we do not know what sort of detector, triggering, or cuts
might be needed to make the discovery, nor which backgrounds we should strive
to better control, nor how much luminosity might be required, and so on.  

At least na\"{i}vely, we can make some progress on these issues by appealing
to the arguments of perturbative unitarity: we know that the loop expansion of
quantum field theory, and hence its predictability, breaks down when couplings
approach values of $4\pi$ or so, and imposing this as an upper bound imposes
an upper limit of $\mathcal{O}(100)$ TeV or so on the possible masses of new
particles~\cite{Altmannshofer:2017yso, DiLuzio:2017chi}. A more refined
analysis of partial wave unitarity shows that the scale of unitarity violation is actually $\sim 80$
TeV~\cite{DiLuzio:2017chi}, and can be even lower in more specific
model-dependent cases. Such unitarity arguments successfully predicted the
appearance of a Higgs boson at the LHC~\cite{Lee:1977eg, Lee:1977yc}, but in
the case of the physics inferred from $b-$decays,
the cut-off scale is too high to form a similar no-lose theorem for
the next generation of colliders.

Here, we attempt to carry out a rather more detailed analysis of the prospects
for discovery of the new physics underlying the $b \rightarrow s \mu^+ \mu^-$
anomalies, at current 
colliders or at proposed future facilities {\em while making as few
  assumptions about the models as possible}. Thus, in the most pessimistic possible scenario, we shall not assume
universal couplings to different generations or minimal flavour violating
couplings; we only include the minimal new physics that explains the $b \rightarrow s
\mu^+ \mu^-$ anomalies whilst
refraining from adding more model-specific structure (that would typically only lower the scale of new physics or make it more easily discoverable). 
It turns out that (at least if one
is prepared to accept a few simple assumptions along the way) one can make
rather detailed and quantitative statements.  
This is possible for a variety of reasons, which we now describe in turn. 

One reason is that, on the theory side, the possible underlying new physics
models are rather limited, at least if one assumes that the new physics
results in an effective low-energy operator coupling a left-handed quark
current to a left-handed leptonic current. This assumption is reasonable not
only because doing so results in a very good fit to the data (in fact, the
best fit to the data, as discussed above), but also because it is highly
plausible theoretically, given that the basic objects in the SM are the
distinct left- and right-handed fermion multiplets. With this
operator, one is limited at tree-level 
to models with either vector leptoquarks (LQs), scalar LQs
or models with new neutral vector particles ($Z^\prime$s) coupling  to
left-handed currents.  

Within this limited range of possible models, there is still a great deal of
room to manoeuvre in terms of choosing couplings. But again we can make
headway by adopting a conservative approach, leading to predictions that are
as pessimistic as possible. For LQ models, for example, it is
perfectly consistent mathematically (although highly unlikely in practice),
that the LQ has only the Yukawa couplings needed to explain the
anomalies, namely to the left-handed lepton doublet containing the muon mass
eigenstate\footnote{The anomalies in the ratios $R_K$ and $R_{K^*}$ could,
  {\em a priori} be due to physics in either muonic or electronic
  operators. But the presence of additional anomalies in purely muonic
  processes \cite{Aaij:2014pli, Aaij:2013qta, Aaij:2015oid, Aaij:2015esa},
  together with the difficulty of accommodating large deviations in flavour
  physics processes involving electrons, both lead us to assume that the new
  physics states couple to muons rather than electrons.} and to the
two left-handed quark doublets containing the $b$- and $s$- quark mass
eigenstates. The presence of any other Yukawa couplings (especially those to
electrons or light quarks) is likely only to increase the discoverability of
the LQ, by providing additional channels for production at a hadron
colliders and additional final states that are relatively easy to
observe.\footnote{There is a danger, {\em e.g.} by adding charm/tau couplings,
  of diluting the LQ decays to clean final states, but to study this
  fully would require an analysis at a level of detail that seems overly
  premature.} 
For $Z^\prime$ models, things are a little more complicated,
because it is not possible to switch on a coupling to $b$ and $s$ quarks
alone: any assignment of charges under the corresponding $U(1)^\prime$ gauge
symmetry to the three quark doublets in the electroweak basis will lead to
other couplings being present in the mass basis.\footnote{Consistency of the
  theory also requires additional particles for the $U(1)^\prime$ gauge
  symmetry to be anomaly-free~\cite{Ellis:2017nrp}.} So we consider two
different conservative models featuring $Z^\prime$ states. In the first model,
we allow only a $bs$ coupling in the mass basis. Though mathematically
inconsistent, strictly speaking, no inconsistencies arise in the collider
phenomenology that we consider here. In the second model, we assume that there
is only a coupling to a single generation of quark and lepton doublets (which
are those that are mostly $b$ and $\mu$, respectively) and assume that all of
the CKM rotation takes place in the down quark sector. Again, for both models
our expectation is that any couplings that are additionally present are likely
to increase discoverability.  

There are also reasons on the experimental side for why a more detailed analysis
of the prospects for discovery at a current or future collider is
possible. Most importantly, we can extrapolate based on the performance of
current colliders, making the conservative assumption that the detector
performance will remain roughly the same.  This extrapolation is simplified by
the fact that the discovery potential of a given machine is largely fixed by
our understanding of the backgrounds. In the particular case of searching for
a narrow resonance in a given channel at a given centre of mass energy, for
example, what is needed is an understanding of the different background
contributions (and their uncertainties) in that channel at that centre-of-mass
(CM) energy. These backgrounds come, of course, from a combination of the
underlying SM physics, which we understand well, together with its
manifestation in the detector, which we assume remains similar to current
detector performance at the extrapolated energies. This extrapolation is
further helped by the fact that the SM is essentially scaleless at the
multi-TeV energies that we consider, so that extrapolation amounts to a simple
re-scaling, using a procedure outlined and validated in
Ref.~\cite{Thamm:2015zwa}: 
in a nutshell, the idea is that the equivalent CM energy at a future collider
that gives the same number of background events as a given CM energy in a
current search will also yield the same upper limit on a putative signal cross
section at that equivalent CM energy.  

Proceeding in this way, we are able to obtain a number of simple results, that
we believe to be robust within our reasonable assumptions. 
We find that a 33 TeV high energy upgrade to the LHC\footnote{Studies to date
  have assumed a 33 TeV centre of mass energy, which we choose as a benchmark, 
but in the future we shall also consider the reduced energy of 27 TeV that can be attained using the 16 T beam magnets currently being designed for FCC-hh.} should be
able to cover most of the $Z^\prime$ parameter space that is under
perturbative control 
in our first model with only $bs$ couplings, while it can cover all of the
parameter space for our second model with CKM-induced couplings to the first
two generations of down-type quarks. A 100 TeV hadron collider has complete
coverage for both models; it can therefore discover or exclude any
perturbative $Z^\prime$ explanation of the anomalies (where the $Z^\prime$ width does not exceed $10 \%$ of its mass). On the LQ side,
considering only pair production via QCD interactions, we find that masses
up to 12 TeV can be ruled out in the scalar case. Limits from single
production are more model-dependent but become important for $\mathcal{O}(1)$
couplings, with sensitivity to LQ masses up to 21 TeV for coupling
values up to $4\pi$.

All of this assumes, of course, that the anomalies currently observed are
really due to new physics. If it turns out that they are not, the exercise
that we have carried out becomes much more academic. But even so, we think
that it gives a useful illustration of the complementarity between indirect
and direct searches and how one can use anomalies that may plausibly arise in
the future, wherever they might occur, to build a concrete
strategy for future colliders and particle physics in general\footnote{For
  some reviews of physics at a 100 TeV hadron collider, see for example
  Refs.~\cite{Arkani-Hamed:2015vfh, Baglio:2015wcg, Golling:2016gvc,
    Contino:2016spe, Mangano:2016jyj}. The indirect sensitivity of future
  lepton colliders has been explored in e.g. Refs.~\cite{Ellis:2015sca,
    Ellis:2017kfi, Durieux:2017rsg, Henning:2014gca, Ge:2016zro,
    Barklow:2017suo}. }.  

The paper is organised as follows: in Section~\ref{sec:Banomaly} we summarise
the effective field theory description of the possible new physics
parameterising the anomalies, justifying our choice of operator, then
describing the possible models that may explain the discrepancy with the
SM. In Section~\ref{sec:projectedsensitivity} we describe the extrapolation
method that we adopt for our study, and present our results. We conclude with
a summary and outlook in Section~\ref{sec:conclusion}.

\section{New physics in $B$ anomalies} 

\label{sec:Banomaly}

\subsection{Effective field theory description} 

Processes involving $b \to s l^+ l^-$ transitions
can be described by a low-energy effective Lagrangian below the weak scale
with the $W^\pm$ boson, $Z$ boson,  Higgs boson and top quark integrated
out.\footnote{If the 
  new physics responsible for the $B$ anomalies is not at low
  energies~\cite{Sala:2017ihs, Ghosh:2017ber, Fuyuto:2015gmk, Datta:2017pfz,
    Bishara:2017pje} then the low-energy effective theory can be matched to
  the SM effective field theory (EFT)~\cite{Alonso:2014csa, Celis:2017doq,
    Celis:2017hod}.} The relevant  
indirect effects of new physics (and SM weak interactions) are encapsulated by
the following four-fermion operators,\footnote{The relation to coefficients of the $\mathcal{O}_{9,10}$ operators in another commonly used basis is given by $c_{9,10} = \pm(c_{LL} \pm c_{LR})/2$~\cite{DAmico:2017mtc}. }  
\begin{align}
\mathcal{L}_\text{eff} &\supset \sum_{l=e,\mu,\tau} \sum_{i=L,R}\sum_{j=L,R} \frac{c^l_{ij}}{\Lambda_{l, ij}^2} \mathcal{O}^l_{ij} \, , \nonumber \\
	&= V_{tb}V^*_{ts}\frac{\alpha_\text{EM}}{4\pi v^2}\sum_{l=e,\mu,\tau} \left( \bar{c}^l_{LL} \mathcal{O}^l_{LL} + \bar{c}^l_{LR} \mathcal{O}^l_{LR} + \bar{c}^l_{RL} \mathcal{O}^l_{RL} + \bar{c}^l_{RR} \mathcal{O}^l_{RR} \right) \, ,
\label{eq:Leff}
\end{align}
where  
\begin{equation}
\mathcal{O}^l_{ij} = (\bar{s}\gamma^\mu P_i b)(\bar{l}\gamma_\mu P_j l) \, .
\end{equation}
In the second line we defined dimensionless Wilson coefficients
$\bar{c}^l_{ij}$ normalised by a conventional factor involving elements of the
CKM matrix $V$ and ratio of the EFT cut-off scale
$\Lambda$ to the weak scale $v 
\simeq 174$ GeV such that 
\begin{equation}
\bar{c}^l_{ij} = \frac{4\pi}{\alpha_\text{EM}
  V_{tb}V^*_{ts}}\frac{v^2}{\Lambda^2}c^l_{ij} \simeq \frac{\left( 36 \text{
      TeV}\right)^2}{\Lambda^2} c^l_{ij} \, . 
\end{equation}
If new particles with couplings to leptons and quarks of size $g_\text{NP}$ are
integrated out at tree-level, 
then $c^l_{ij} \sim \mathcal{O}(g_\text{NP}^2)$ and since,
  according to our criterion, the limit of
validity of perturbative 
unitarity is reached when $g_\text{NP} \sim 4\pi$, this sets an approximate
upper limit on the cut-off scale\footnote{The perturbativity condition is
  sometimes also taken to be $g_\text{NP}^2 \sim 4\pi$~\cite{Altmannshofer:2017yso}, in which case the
  cut-off is $\Lambda_\text{max} \sim 127 \text{ TeV} / \sqrt{\bar{c}^{l}_{ij}} \sim 110 \text{ TeV}$.} of
\begin{equation}
\Lambda_\text{max} \sim \frac{450 \text{ TeV}}{\sqrt{\bar{c}^l_{ij}}} \, .
\end{equation}
For example, with $|\bar{c}^l_{ij}| \simeq 1.33$, as found in certain best fit
values~\cite{DAmico:2017mtc}, we have $\Lambda_\text{max}  \lesssim 390$
TeV. A more detailed analysis of partial wave unitarity yields a $80$ TeV bound~\cite{DiLuzio:2017chi}. However, 
other experimental and theoretical bounds will lead to a more restrictive
upper limit on the scale of new physics, as we discuss below. 

Many global fits to the flavour anomalies have been performed e.g.~\cite{Descotes-Genon:2013wba, Altmannshofer:2013foa, Beaujean:2013soa, Hurth:2013ssa, Capdevila:2017bsm, Altmannshofer:2017yso, Hiller:2017bzc, Geng:2017svp, Ciuchini:2017mik, DAmico:2017mtc}. Ref.~\cite{DAmico:2017mtc}, for example, finds that an individual fit to one operator at a
time in the muonic sector favours $\bar{c}^\mu_{LL} \simeq -1.33$ at $>
4\sigma$ significance. A similar conclusion holds for a global fit allowing
several 
operators to vary simultaneously, which then allows an additional sub-dominant
contribution from $\bar{c}^\mu_{LR}$ (though $\bar{c}^\mu_{LR}$ alone cannot
explain the anomalies since it predicts the pattern $R_{K^*} > 1$ when $R_K <
1$ or vice versa). The coefficients $\bar{c}^\mu_{RR}$ and $\bar{c}^\mu_{RL}$,
whose contributions must be large to have an effect since their SM
interference terms are suppressed, are disfavoured by the relative directions
of their pulls on 
$R_K$ and $R_{K^{*}}$.  

In individual fits to $R_{K^{(*)}}$ for electronic operators, the anomalies are
also well described by either $\bar{c}^e_{LL}, \bar{c}^e_{LR}$, or
$\bar{c}^e_{RR}$ (though the latter two require larger coefficient values due
to their suppressed SM interference). Nevertheless the significance decreases
substantially in a global fit including other observables, which shows
a clear preference for non SM contributions in decays to muons rather than in
decays to electrons\footnote{Ref.~\cite{Ghosh:2014awa} first pointed out an indication of lepton flavour universality violation from a global fit, though more data is needed to conclusively establish this~\cite{Hurth:2017hxg}.}. We shall 
therefore assume new physics to reside solely in the muonic sector and in
$\bar{c}^\mu_{LL}$ in particular. This restricts the type of heavy particles that can be
integrated out to give ${\bar c}^\mu_{LL}$ in the EFT, as we discuss next.

\subsection{$Z^\prime$ and LQ models to explain the discrepancy}
\label{sec:ZprimeLQ}
\begin{figure}
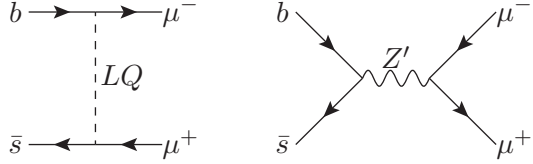

\begin{center}
\begin{axopicture}(150,50)(0,0)
% Leptoquark diagram
\Line[arrow](50,0)(25,0)
\Line[arrow](25,0)(0,0)
\Line[arrow](0,50)(25,50)
\Line[arrow](25,50)(50,50)
\Line[dash](25,50)(25,0)
\Text(-5,50)[c]{$b$}
\Text(-5,0)[c]{$\bar s$}
\Text(57,50)[c]{$\mu^-$}
\Text(57,0)[c]{$\mu^+$}
\Text(35,25)[c]{$LQ$}

% Z' diagram
\Line[arrow](100,50)(125,25)
\Line[arrow](125,25)(100,0)
\Line[arrow](175,50)(150,25)
\Line[arrow](150,25)(175,0)
\Photon(125,25)(150,25){3}{3}
\Text(137.5,33)[c]{$Z^\prime$}
\Text(95,50)[c]{$b$}
\Text(95,0)[c]{$\bar s$}
\Text(182,50)[c]{$\mu^-$}
\Text(182,0)[c]{$\mu^+$}
\end{axopicture}
\end{center}
\caption{Feynman diagrams of the two tree-level possibilities for mediating
  an effective operator 
that explains discrepancies in $B \to K^{(*)} \mu^+ \mu^-$ decays as compared
to SM predictions. The
diagram on the left hand side shows mediation by a scalar, whereas the
right-hand side shows mediation by a flavour
dependent $Z^\prime$. \label{fig:feyn}}
\end{figure}

At tree level there are only a few candidates to consider for mediating 
the interactions responsible for the $B$ anomalies. These are so-called LQs,
that can be either scalar or vector, and $Z^\prime$ vector bosons. We shall
assume that in each scenario, the new fields are {\em unique}\/
representations of the Lorentz group and the SM, i.e.\ we are not
considering multiple identical fields. 
Feynman
diagrams for the relevant interactions are shown in Fig.~\ref{fig:feyn}.
When the mass of the LQ or $Z^\prime$ is much larger than the mass of
the decaying $B$ meson, matching to the effective field theory in
Eq.~\ref{eq:Leff} should provide an accurate approximation to order $m_B /
\Lambda$, where $\Lambda$ is the mass of the LQ or $Z'$. 

Other explanations for the anomalies arise at the loop level. In this case, in order to explain the required size of the non-standard contributions to $B
\to K^{(*)} \mu^+ \mu^-$ decays, the new particles mediating the interaction
must be relatively light and so are more easily discoverable; we therefore restrict our attention to the more
conservative case of heavier tree-level induced new physics.   

The preference of fits for the $\mathcal{O}_{LL}^\mu$ operator picks out
particular combinations of quantum numbers allowed for the
LQs~\cite{DAmico:2017mtc, Hiller:2017bzc, Capdevila:2017bsm} . For 
the scalar case this is the triplet LQ $S_3$, with quantum numbers
$(\bar{3}, 3, \frac{1}{3})$ under $SU(3)_c \times SU(2)_L \times U(1)_Y$,
whose Yukawa couplings to the quark and lepton doublets $Q$ and $L$ are of the form
\begin{equation}
y_3 QL S_3 + y_q QQ S_3^\dagger + \text{h.c.} \, .
\end{equation}
The term proportional to $y_q$ induces proton decay and is typically set to
zero by imposing baryon number conservation. For the vector case, the
$\mathcal{O}_{LL}$ operator may be generated by integrating out a singlet
$V_1$ or a triplet $V_3$ with quantum numbers $(\bar{3}, 1, \frac{2}{3})$ and
$(3, 3, \frac{2}{3})$, respectively. The possible couplings are 
\begin{equation}
y_3^\prime V_3^\mu \bar{Q}\gamma_\mu L + y_1 V_1^\mu \bar{Q}\gamma_\mu L  + y_1^\prime V_1^\mu \bar{d}\gamma_\mu l + \text{h.c.} \, .
\end{equation}
We focus on the couplings generating our operator of interest,
$\mathcal{O}_{LL}^\mu$. Integrating out the LQs with mass $M$ and coupling $y$
gives the Wilson coefficient~\cite{Hiller:2017bzc} 
\begin{equation}
\bar{c}_{LL}^\mu = \kappa \frac{4\pi v^2}{\alpha_\text{EM} V_{tb} V^*_{ts}} \frac{|y|^2}{M^2} \, ,
\label{eq:cbarLLfromLQ}
\end{equation}
where $\kappa = 1, -1, -1$ and $y=y_3,y_1,y_3'$ for $S_3, V_1, V_3$, respectively. 

For $Z^\prime$ vector bosons, the minimal Lagrangian containing the couplings
responsible for generating $\mathcal{O}_{LL}^\mu$ at low energy is given
by~\cite{Allanach:2015gkd,DAmico:2017mtc}. 
\begin{equation}
\mathcal{L}_{Z^\prime}^\text{min.} \supset \left( g_L^{sb} Z^\prime_\rho \bar{s}\gamma^\rho P_L b  + \text{h.c.} \right) + g_L^{\mu\mu} Z^\prime_\rho \bar{\mu} \gamma^\rho P_L \mu \, ,
\label{eq:minimalZprime}
\end{equation}
which contributes to the $\mathcal{O}_{LL}^\mu$ coefficient with
\begin{equation}
\bar{c}_{LL}^\mu =  -\frac{4\pi v^2}{\alpha_\text{EM} V_{tb} V^*_{ts}} \frac{g_L^{sb} g_L^{\mu\mu}}{M_{Z^\prime}^2} \, .
\end{equation}

Couplings to some other SM fermions are required by $SU(2)_L$
invariance and some additional couplings to other flavours of quark are necessarily generated by CKM
rotations when going from the 
weak to the mass eigenbasis. However, given that these additional interactions are
more model-dependent than the ones we write above, we shall take the Lagrangian of
Eq.~\ref{eq:minimalZprime} as our minimal model (which we call the na\"{i}ve
$Z^\prime$ model). Although strictly, the model  is incomplete without the
additional couplings, 
the na\"{i}ve $Z^\prime$ model is the most conservative
possible case to study; additional couplings will only raise the
$Z^\prime$ production 
cross-section, by including couplings to the first two quark generations, and
increase the total decay width which is in tension with other constraints. 
Hence, if a future collider covers some portion of the viable parameter space of
the na\"{i}ve $Z^\prime$ model, then we know that a more realistic and complete
model will also be covered there (and then some).

To illustrate the size of such effects in a more complete model we shall also
consider the case where 
the $Z^\prime$ couples only to third generation left-handed quarks and
left-handed muons and neutrinos in the weak
basis. The 
couplings to the first two generations of quarks then arise from CKM
rotations, which we assume 
to be entirely in the down sector. Additionally, if we assume that in the
weak eigenbasis all left-handed
lepton mixing resides in the neutrino sector, we have a logically consistent
model which contains only a coupling to left-handed muons and some family
mixture of neutrinos. The precise family mixture of neutrinos is immaterial for
collider experiments, since each neutrino is essentially massless and leaves
an identical missing momentum signature in detectors. The relevant interaction
terms in the Lagrangian for this `$33\mu\mu$' model are given by   
\begin{align}
\mathcal{L}_{Z^\prime}^{33\mu\mu} &\supset g^q_L Z^\prime_\rho \left[ \bar{t}\gamma^\rho P_L t + |V_{tb}|^2 \bar{b} \gamma^\rho P_L b + |V_{td}|^2\bar{d}\gamma^\rho P_L d + |V_{ts}|^2\bar{s}\gamma^\rho P_L s \right. \nonumber \\
 &\left. + \left( V_{tb}^* V_{ts} \bar{b}\gamma^\rho P_L s + V_{ts}^* V_{td}
   \bar{d}\gamma^\rho P_L s + \text{h.c.} \right) +
   g_L^{\mu\mu}\left(\bar{\mu}\gamma^\rho P_L \mu +  
   \sum_i \bar{\nu}_i U_{i\mu} \gamma^\rho P_L U_{\mu i}^*\nu_i \right) \right],
\label{eq:3rdgenZprimeLagrangian}
\end{align}
where $U$ denotes the PMNS matrix involved in lepton mixing.

With these LQ and $Z^\prime$ models in hand, we now turn to their discovery prospects. Some previous studies have examined the 13 TeV LHC's ability to
discover other 
effects 
caused by new physics involved in the errant
$b-$decays assuming any mediator is not too heavy. In
Ref.~\cite{Allanach:2015gkd}, LHC bounds on $Z^\prime$ 
models that explain the $b-$anomalies from di-muon resonances were placed,
{assuming a universal $Z^\prime$ coupling to the first two generations of
  quarks}. Ref.~\cite{Buttazzo:2017ixm} also examined current LHC constraints
on LQs
and performed na\"{i}ve re-scaling to estimate the sensitivity at higher
luminosities in models which explain both the $b \rightarrow s \mu^+ \mu^-$
anomalies that we consider and  
additional ones inferred in $b \rightarrow c \tau \bar \nu_\tau$ decays\footnote{In the
present 
paper, we do not
consider physics due to these charged current decays because their SM
predictions are subject 
to larger theoretical errors. Moreover, the size of those effects requires a low mass scale that would make the new physics responsible more easily discoverable than the source of the neutral current $b$-anomalies.}.
Ref.~\cite{Dorsner:2017ufx} also examined the LHC's ability to detect scalar LQs of 1 TeV
mass of the 
type that we shall examine.
Ref.~\cite{Greljo:2017vvb} 
examines the 
di-lepton final state for effective field theory operators caused by LQs or
$Z^\prime$s. Some sensitivity is found under the assumption of minimal flavour
violation for light enough $Z^\prime$s. 

In our study we look towards future colliders at higher luminosity and energy. In the next Section we shall estimate the projected limits on the $Z^\prime$ and $LQ$ masses in our conservative models by extrapolating from direct searches at the LHC.

\section{Projected sensitivity on $Z^\prime$ and LQs}
\label{sec:projectedsensitivity}

\subsection{Limit extrapolation method} 
\label{sec:extrapolationmethod}

We follow the approach of Ref.~\cite{Thamm:2015zwa} to extrapolate the limits
on direct searches for new resonances at the LHC to higher energy and
luminosity. The method assumes that such a limit is entirely driven by the
number of background events, so that finding the equivalent mass at a future
collider that gives the same background as a given mass in a current search
will also yield the same upper limit on a putative signal cross section at that
equivalent mass.   

Concretely, the background cross-section at a resonance
mass $M$ and centre of mass collision energy $\sqrt{s}$ is 
\begin{equation}
\sigma_B(M, s) \propto \sum_{i, j} \int_{M^2-\Delta \hat{s}}^{M^2+\Delta\hat{s}} d\hat{s} \frac{dL_{ij}}{d\hat{s}} \hat{\sigma}_{ij}(\hat{s}) \, ,
\end{equation}
where $\hat \sigma_{ij}(\hat s)$ is the partonic cross section for production
of the resonance by partons $i$ and $j$ evaluated at a partonic centre of mass
energy $\sqrt{\hat s}$
and the parton luminosity function $dL_{ij} / d\hat{s}$ for the initial state parton pair labelled by $i$ and $j$ is given by
\begin{equation}
\frac{dL_{ij}}{d\hat{s}} = \frac{1}{s} \int_{\hat{s}/s}^1 \frac{dx}{x} f_i \left(x, \mu^2 \right) f_j \left(\frac{\hat{s}}{s x}, \mu^2 \right).
\end{equation}
We set the factorisation scale $\mu = \sqrt{\hat{s}}$. We assume that the
resonance is sufficiently narrow, $\Delta \hat{s} \ll M^2$, such that the
partonic luminosity is approximately constant in the integration region. Since
the background consists of SM processes at energies far above the weak scale,
the partonic cross-section should scale like $\hat{\sigma}_{ij} \propto
1/\hat{s}$. The total background cross-section then simplifies to  
\begin{equation}
\sigma_B(M,s) \propto \frac{\Delta \hat{s}}{M^2} \sum_{i,j} C_{ij} \frac{dL_{ij}}{d\hat{s}}(M, s) \, ,
\end{equation}
where $C_{ij} = \hat{s} \hat{\sigma}_{ij}$ is approximately constant. The
number of background events at a given luminosity $L$ is $N_B = L \cdot
\sigma_B(M,s)$. If a $95 \%$ confidence level (CL) limit on a signal
cross-section is set for a given resonance mass $M_0$ at a present collider
(with energy $\sqrt{s_0}$ and luminosity $L_0$), then we find the equivalent
mass $M^\prime$ for which the limit applies
 at a future collider (with energy
$\sqrt{s^\prime}$ and luminosity $L^\prime$) by the assumption that the same
limit is applicable when $N_B^\prime=N_B^0$, i.e.\
\begin{equation}
L_0 \cdot \sum_{i,j} C_{ij} \frac{dL_{ij}}{d\hat{s}}(M_0, s_0) = L^\prime \cdot \sum_{i,j} C_{ij} \frac{dL_{ij}}{d\hat{s}}(M^\prime, s^\prime) \, .
\end{equation} 
The fixed relative width $\Delta \hat{s} / M^2$ and other prefactors have
cancelled out, leaving a straightforward equation to solve for $M^\prime$. The
constants $C_{ij}$ can be normalised such that they represent the relative
weights of the contributions from each parton pair. 

This method introduces some arbitrariness in the starting point of the extrapolated exclusion curve, since it depends on a re-scaling by the luminosity ratio $L_0 / L^\prime$. If $L^\prime = L_0$ then the smallest mass ${M_0}_\text{min}$ at the lower end of the current collider sensitivity will be extrapolated to the starting point ${M^\prime}_\text{min}$ of the exclusion curve at the future collider. On the other hand if $L^\prime > L_0$ then the starting point will be at a higher mass point, while $L^\prime < L_0$ would reach lower masses. A conservative procedure to account for this artificial effect is to smoothly vary the future collider luminosity up to $L^\prime$ during the extrapolation and take the strongest limit for each mass point, which only affects the limit for masses below $M^\prime_\text{min}$, and in any case is more conservative than a realistic limit~\cite{Thamm:2015zwa}.

This extrapolation method has been validated against a cut-and-count-based
analysis for di-lepton searches in Ref.~\cite{Thamm:2015zwa}, where agreement
is found up to a factor of two for a width of $\Delta\hat{s} / M^2 =
10\%$. Results from the approximate method outlined here can then be trusted
in so far as a more complete
analysis does not give limits too far off from a cut-and-count-based
one. While more realistic experimental analyses will certainly use more
refined methods that go beyond our assumptions, the approximation is
sufficient for a rough estimate of future collider sensitivity and should help
motivate
a more detailed study.

\subsection{$Z^\prime$ sensitivity} 

\begin{figure}
\centering
\includegraphics[width=0.49 \textwidth]{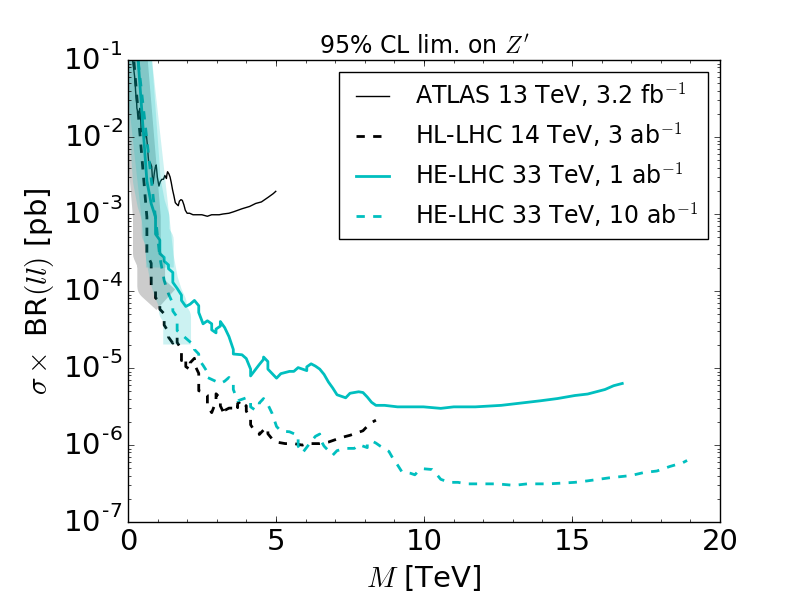}
\includegraphics[width=0.49 \textwidth]{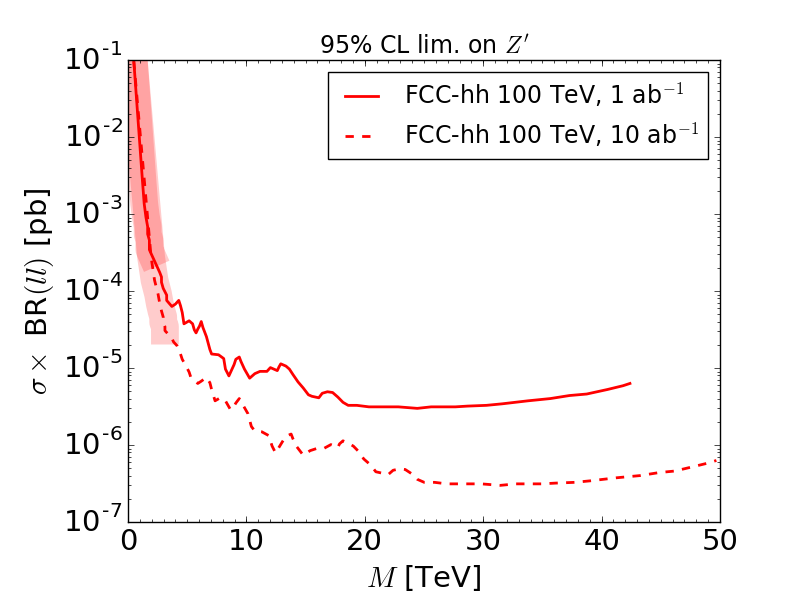}
\caption{Projected sensitivity of di-muon resonance searches of some future
  hadron colliders to $Z^\prime$ models that may explain anomalous $B \to K^{(*)}
  \mu^+ \mu^-$ decay results for the luminosities and centre of mass energies
  given in the legend. Shaded parts of the curve indicate the conservative extrapolation method at low masses that underestimates the actual limit. }
\label{fig:minimalZprimeexclusion}
\end{figure}

We extrapolate limits from the ATLAS 13 TeV search in the di-muon final state
at $\sqrt{s} = 13$ TeV and 3.2 fb$^{-1}$~\cite{Aaboud:2016cth}\footnote{We used the obtained LHC limit rather than the expected sensitivity. However, since the
  limit and the sensitivity are close (within about $2\sigma$), this is a
  reasonable approximation.}. The dominant
backgrounds come from Drell-Yan, $t\bar{t}$ and di-boson production. Using the
procedure described in Section~\ref{sec:extrapolationmethod}, we obtain the
projected limits displayed in Fig.~\ref{fig:minimalZprimeexclusion}. The solid
black line in the left plot is the current $95\%$ CL limit from the ATLAS 13
TeV analysis. In dashed black is the projected limit for HL-LHC at 14 TeV with
3 ab$^{-1}$, while the solid and dashed lines in cyan are for the HE-LHC at 33
TeV with 1 and 10 ab$^{-1}$, respectively. The plot on the right shows the
corresponding FCC-hh 100 TeV limits in solid (dashed) red for 1 (10)
ab$^{-1}$. The shaded regions on the curves indicate the point at which the extrapolation method underestimates the actual limit at low masses, as explained in Section~\ref{sec:extrapolationmethod}.

One may note various features in Fig.~\ref{fig:minimalZprimeexclusion} that might seem surprising {\em prima facie}: for
example, it appears that the 14 TeV  3 ab$^{-1}$ HL-LHC can reach lower in
$\sigma \times BR$ than 
the 10 ab$^{-1}$ 33 TeV HE-LHC for $M< 6$ TeV. This is caused by the behaviour
of regions dominated by high backgrounds at
lower masses: if one increases the centre of
mass energy from the LHC to higher collider energies then 
this background-dominated region will correspondingly move to higher masses. On the other end we see
that at the highest values of $M$ the HE-LHC is the most sensitive, as
expected. While these sensitivity limits are purely a function of the background, the actual limit set for a given $Z^\prime$ mass and coupling also depends on the signal cross-section, which is larger at higher collider energies. Therefore a lower-energy collider whose limit curve reaches further down than that of a higher-energy collider does not necessarily translate to better sensitivity in a model's parameter space. 

The actual $Z^\prime$ mass that can be excluded for a $B$-anomaly-compatible model depends on the specific couplings of the $Z^\prime$ and its total decay width. We calculated the Drell-Yan cross-section for $p p \to Z^\prime \to \mu^+ \mu^-$ as a function of these couplings using the following expression in the narrow width approximation, cross-checked with {\tt MadGraph}~\cite{Alwall:2014hca},
\begin{equation}
\sigma_{p p \to Z^\prime \to \mu^+ \mu^-} = 16\pi^2 \sum_{i,j} \left( \frac{S_{Z^\prime}}{S_i S_j} \frac{C_{Z^\prime}}{C_i C_j} \frac{\Gamma_{Z^\prime \to \bar{q}_i q_j}}{M_{Z^\prime}} \frac{1}{s} \left. \frac{dL_{ij}}{d\tau}\right|_{\tau=M_{Z^\prime}^2/s} \right) \text{BR}(Z^\prime \to \mu^+ \mu^-) \, ,
\end{equation}
where $S_i$ and $C_i$ are the number of spin and colour degrees of freedom
of parton $i$
respectively, and the parton luminosity function is 
\begin{equation}
\frac{dL_{ij}}{d\tau} = \int_{\tau}^1 \frac{dx}{x} f_i \left(x, \mu^2 \right) f_j \left(\frac{\tau}{x}, \mu^2 \right) \, .
\end{equation}
The decay rate for $Z^\prime$ into fermions with coupling $g_f$, assuming $m_f \ll M_{Z^\prime}$, is given by 
\begin{equation}
\Gamma_{Z^\prime \to \bar{f}_i f_j} = \frac{C}{24\pi} g_f^2 M_{Z^\prime}^2 \, .
\end{equation}
For the parton distribution functions $f(x,\mu^2)$ we use the 5-flavour
NNPDF2.3LO ($\alpha_s(M_Z)=0.119$) set~\cite{Ball:2012cx} with {\tt LHAPDF}~\cite{Buckley:2014ana} and
fix the factorisation scale to be $M_{Z^\prime}^2$. We consider $b$ quarks to
be in the initial PDFs of the proton, thus re-summing large logarithms on the
initial $b$ quark line~\cite{Lim:2016wjo}. The Feynman diagram for hadron
collider production is therefore identical to the right-hand plot of
Fig.~\ref{fig:feyn}.

\begin{figure}
\centering
\includegraphics[width=0.49 \textwidth]{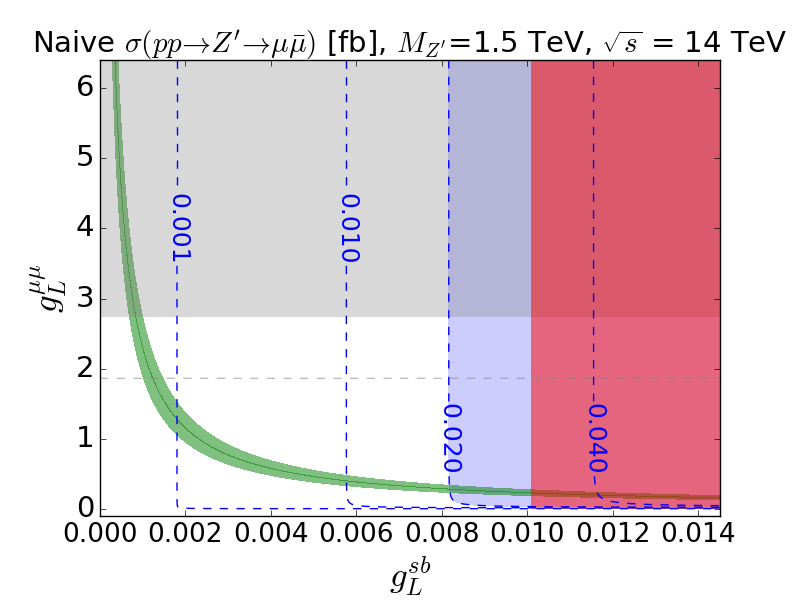}
\includegraphics[width=0.49 \textwidth]{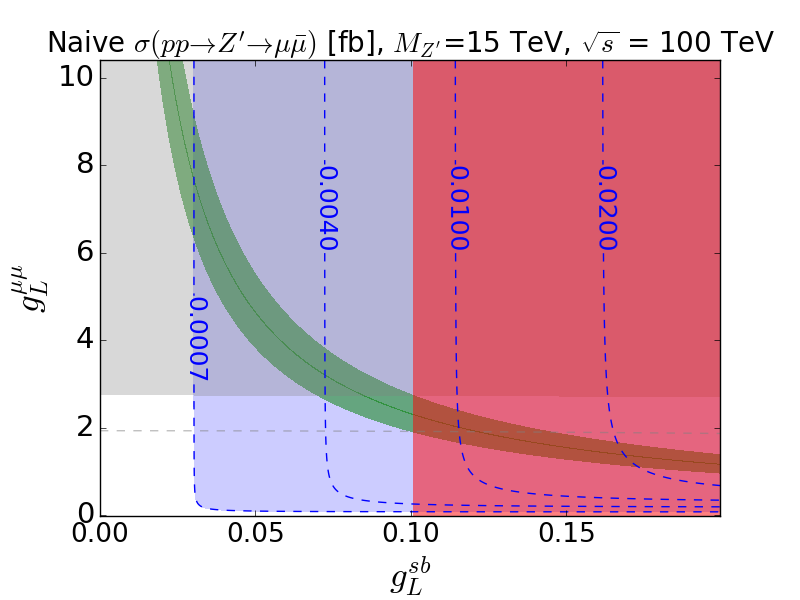}
\includegraphics[width=0.49 \textwidth]{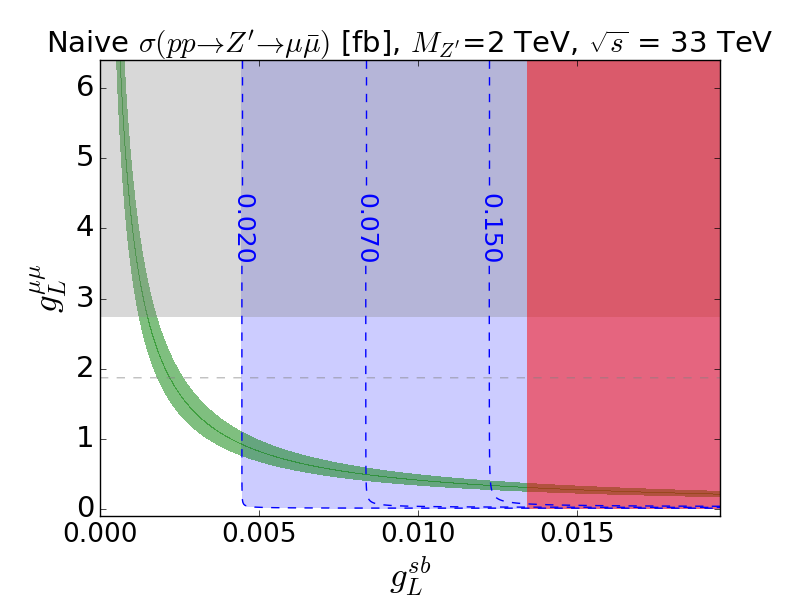}
\includegraphics[width=0.49 \textwidth]{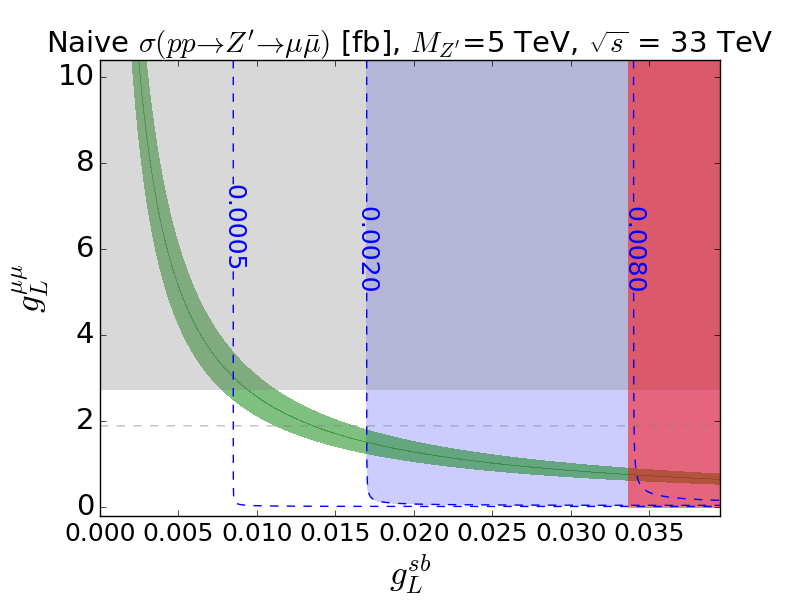}
\caption{Parameter space of $Z^\prime$ models that explain $B
  \to  K^{(*)} \mu^+ \mu^-$ decay results for the na\"{i}ve $Z^\prime$ model for
  different future colliders and $M_{Z^\prime}$ assumptions. The
  horizontal (grey) shaded 
  region violates the narrow width approximation. The vertical (red) region
  extending to the right hand side of each plot shows the limit coming from
  $B_s - \bar B_s$ mixing measurements. The (green) curve displays the region
  that   fits $B   \to  K^{(*)} \mu^+ \mu^-$ decay results. Above the dashed
  (grey) horizontal line, the coupling reaches a Landau pole below the Planck
  scale. The darker dashed (blue) contours are labelled with the expected
  production cross-section times branching ratio in fb. The shaded (blue)
  region shows the expected sensitivity at the future collider from 
  di-muon resonance searches derived from
  Fig.~\ref{fig:minimalZprimeexclusion}.}
\label{fig:minimalZprimecouplingsplot}
\end{figure}

Using these expressions and the extrapolated limits of
Fig.~\ref{fig:minimalZprimeexclusion}, the resulting parameter space for the
na\"{i}ve model is shown in Fig.~\ref{fig:minimalZprimecouplingsplot}. As
discussed in Section~\ref{sec:ZprimeLQ}, we take the na\"{i}ve model of
Eq.~\ref{eq:minimalZprime} defined by only a $Z^\prime$ coupling to $\bar{b} s
+ \bar{s}b$ and $\mu^+ \mu^-$, and nothing else, as the most conservative
possible case. While other couplings should necessarily be present, the
na\"{i}ve $Z^\prime$ serves as a useful scenario to assess the sensitivity of
a future collider since any model that seeks to explain the $B$-anomalies must
have at least these couplings, while other interactions are more
model-dependent.

The line and colour coding for Fig.~\ref{fig:minimalZprimecouplingsplot} is as
follows: the blue-shaded region covering the area vertically towards the right
corresponds to the extrapolated $95\%$ CL limit for the highest luminosity at
the collider energy shown in the plot title; the grey-shaded region excluding
the area horizontally 
towards the top is where the $Z^\prime$ width exceeds $10\%$; the vertical red
region is excluded by too large a contribution to $B_s - \bar{B}_s$ mixing
which constrains $|\bar{g}_L^{sb}| \lesssim \sqrt{2}M_{Z^\prime} / (210
\text{ TeV})$~\cite{DAmico:2017mtc}; the green region is compatible with the
$B$ anomaly within $1\sigma$ of the best fit value of
Ref.~\cite{DAmico:2017mtc}; the blue (mostly) vertical dashed lines are the contours of
cross-section in units of femtobarns; and the horizontal grey dashed line is
where~\cite{Allanach:2015gkd} 
\begin{equation}
\frac{\Gamma_{Z^\prime}}{M_{Z^\prime}} \lesssim \frac{\pi}{2}\frac{1}{\ln(M_{pl}/M_{Z^\prime})} \, ,
\end{equation}
indicating that the $Z^\prime$ couplings will hit a Landau pole before the
Planck scale; the region above this dashed line is
therefore theoretically disfavoured. This last condition is model dependent as
the Feynman diagram 
contributing to the decay width, given by the imaginary part of the $Z^\prime$
propagator, will also contribute to the renormalisation group running from the
real part of the propagator. While this perturbativity condition is weakened
by new vector bosons contributing to the running, it is strengthened by
the addition of scalars or fermions.  

\begin{figure}
\centering
\includegraphics[width=0.49 \textwidth]{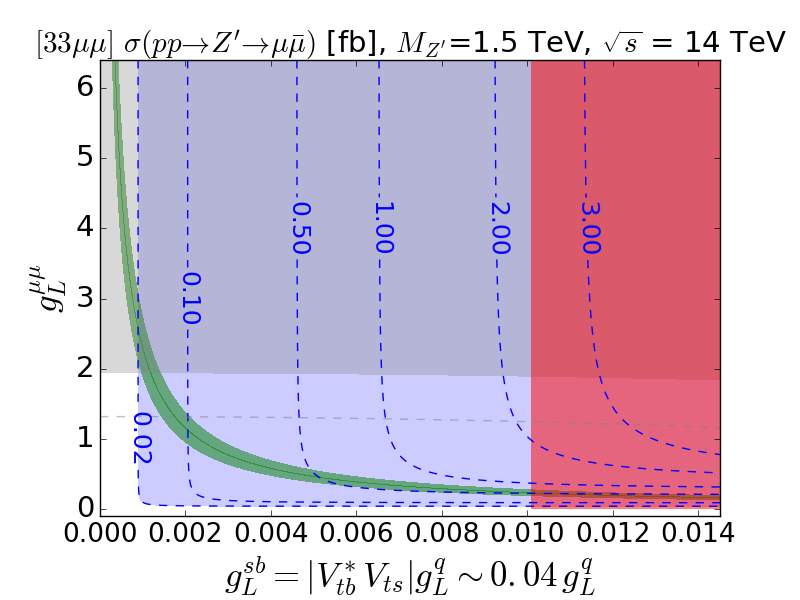}
\includegraphics[width=0.49 \textwidth]{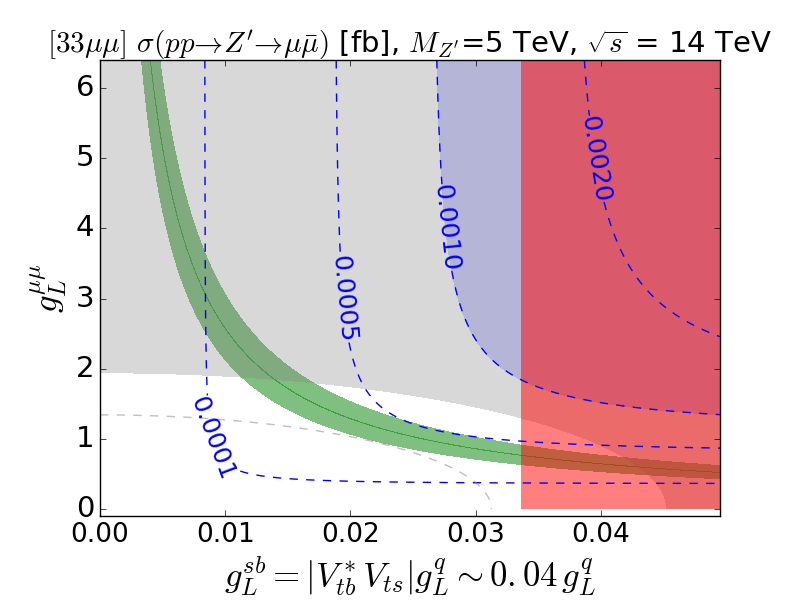}
\includegraphics[width=0.49 \textwidth]{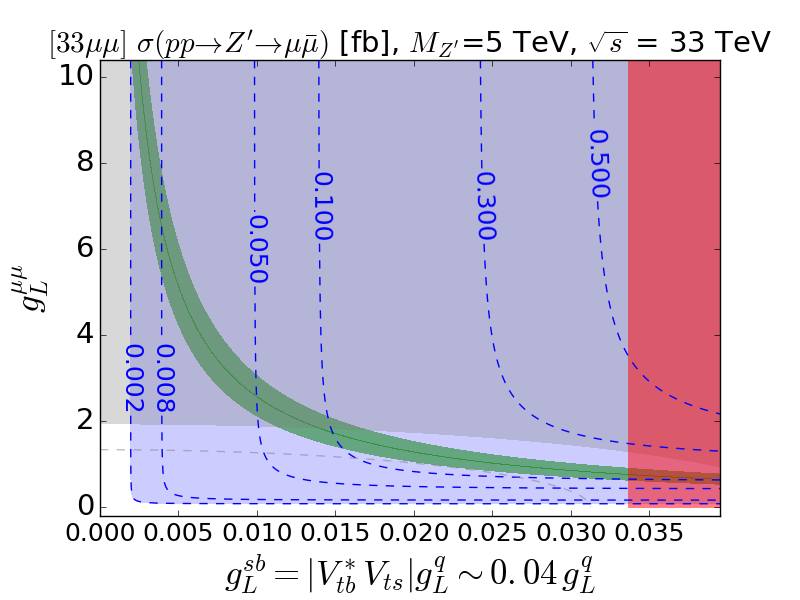}
\includegraphics[width=0.49 \textwidth]{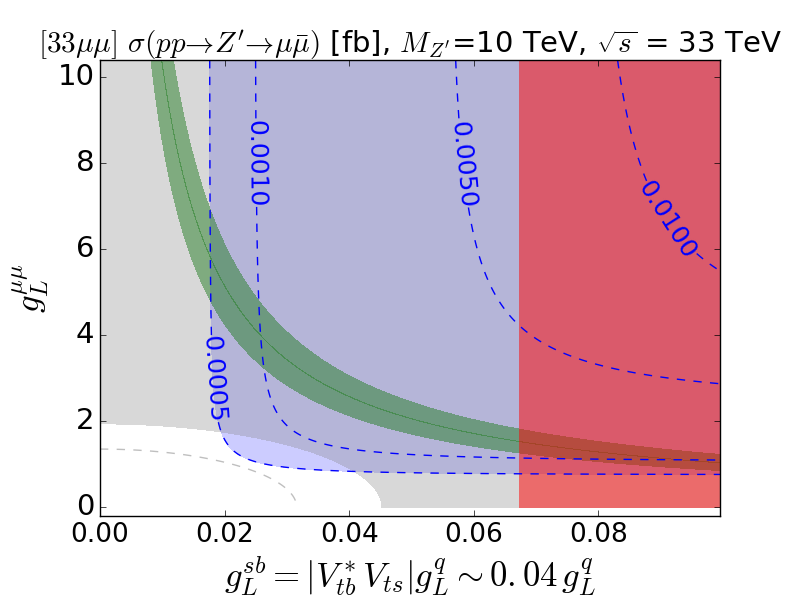}
\caption{Parameter space of $Z^\prime$ models that explain $B
  \to  K^{(*)} \mu^+ \mu^-$ decay results for the 33$\mu\mu$ model for
  different future colliders and $M_{Z^\prime}$ assumptions. The
  horizontal (grey) shaded 
  region violates the narrow width approximation. The vertical (red) region
  extending to the right hand side of each plot shows the limit coming from
  $B_s - \bar B_s$ mixing measurements. The (green) curve displays the region
  that   fits $B   \to  K^{(*)} \mu^+ \mu^-$ decay results. Above the dashed
  (grey) horizontal line, the coupling reaches a Landau pole below the Planck
  scale. The darker dashed (blue) contours are labelled with the expected
  production cross-section times branching ratio in fb. The shaded (blue)
  region shows the expected sensitivity at the future collider from 
  di-muon resonance searches derived from
  Fig.~\ref{fig:minimalZprimeexclusion}.}
\label{fig:3rdgenZprimecouplingsplot}
\end{figure}

The top left plot in Fig.~\ref{fig:minimalZprimecouplingsplot} indicates that
the HL-LHC at 14 TeV and 3 ab$^{-1}$ is barely sensitive to a na\"{i}ve
$Z^\prime$ when its mass is $1.5$ TeV. This may seem low but we recall that in
the na\"{i}ve model the only production mechanism for $Z^\prime$ in Drell-Yan
is through $b$ and $s$ initial state partons~\footnote{For a study of other
  possible production mechanisms with these couplings, see
  Ref.~\cite{Dalchenko:2017shg}.}. On the other extreme end of collider reach
is the FCC-hh at 100 TeV, shown on the top right for $M_{Z^\prime} = 15$
TeV. We see that a 15 TeV $Z^\prime$ is at the limit of being
anomaly-compatible and evading the constraints from both $B_s - \bar{B}_s$
mixing and Landau poles. Nevertheless, the blue region corresponding to FCC-hh
with 10 ab$^{-1}$ can easily cover all of the parameter space of interest. For
lower luminosities the sensitivity can be read off from the cross-section
contours and the corresponding limits in
Fig.~\ref{fig:minimalZprimeexclusion}.   

Between the CM energies of HL-LHC and FCC-hh is the 
HE-LHC at 33 TeV CM energy,
displayed in the bottom row of Fig.~\ref{fig:minimalZprimecouplingsplot} for a
$Z^\prime$ mass of 2 TeV on the left and 5 TeV on the right. The sensitivity
drops off such that the HE-LHC no longer covers any non-excluded parameter
space for $M_{Z^\prime} \gtrsim 7$ TeV.

\begin{figure}
\centering
\includegraphics[width=0.49 \textwidth]{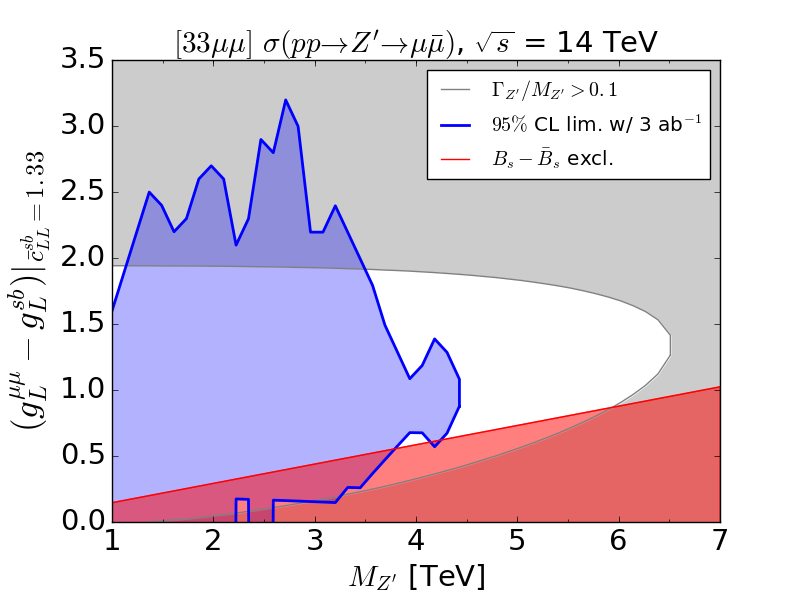}
\includegraphics[width=0.49 \textwidth]{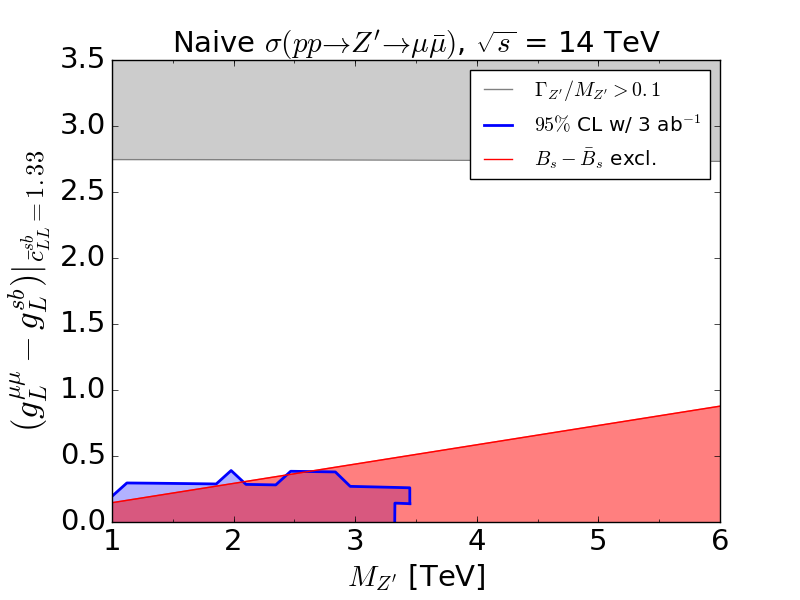}
\includegraphics[width=0.49 \textwidth]{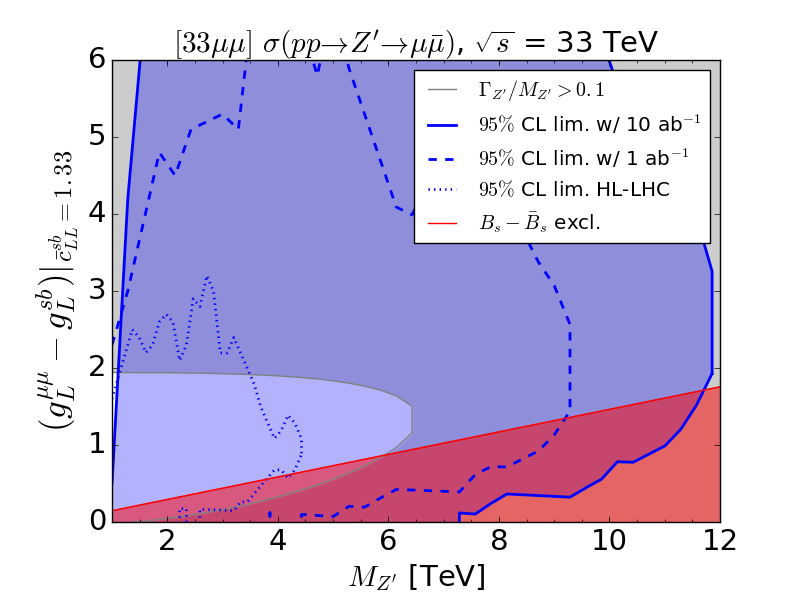}
\includegraphics[width=0.49 \textwidth]{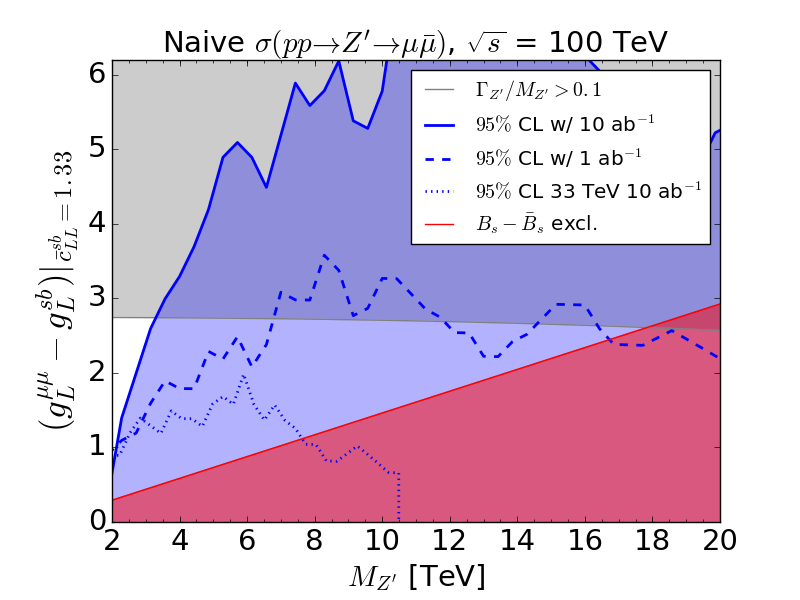}
\caption{\label{fig:zprimesum} Summary of b-anomaly explaining $Z^\prime$
  search sensitivity in the 
mass-coupling plane for various different future hadron collider options in
the na\"{i}ve model and the $33\mu\mu$ model. The blue shaded region shows the
expected 
sensitivity at the future collider from  
  di-muon resonance searches derived from
  Fig.~\ref{fig:minimalZprimeexclusion}. The red region
  extending to the right hand side of each plot shows the limit coming from
  $B_s - \bar B_s$ mixing measurements.
The
  grey shaded 
  region violates the narrow width approximation.} 
\end{figure}

To illustrate the possible sensitivity to a more realistic model, in
Fig.~\ref{fig:3rdgenZprimecouplingsplot} we show the reach for the 
$33\mu\mu$ $Z^\prime$ model defined by the Lagrangian of
Eq.~\ref{eq:3rdgenZprimeLagrangian}. There, 
the couplings to third-generation left-handed quarks induce a coupling to the
first two generations of quarks through the CKM matrix. This raises the production
cross-section through the additional initial state partonic channels, and also
increases the total decay width. In the top left-hand plot we see that a 1.5 TeV
$Z^\prime$ is now accessible to the HL-LHC in all of its favoured parameter
space, with the top right-hand plot indicating that the new limit of sensitivity of
the HL-LHC for this more realistic model is raised to $M_{Z^\prime} \lesssim
4$ TeV. From the bottom two plots, with $M_{Z^\prime} = 5$ (10) TeV on the
left (right), we conclude that the 33 TeV HE-LHC at its highest luminosity can
cover all the parameter space of interest for all favoured masses. Indeed, we
see that for $M_{Z^\prime} \gtrsim 10$ TeV the anomaly-compatible region lies
entirely within the grey and red areas and yet is still covered by the blue-shaded
area. The FCC-hh with even more energy will therefore also be sensitive to the
entire mass range, so we omit its plot.

To summarise the projected reach, we now study the behaviour of the bounds and future collider coverage of
$Z^\prime$ models shown in Figs.~\ref{fig:minimalZprimecouplingsplot} and
\ref{fig:3rdgenZprimecouplingsplot} for a continuously varying $M_{Z^\prime}$,
shown on the abscissa. 
We scan along 
the central green line in those figures, corresponding to the central
inferred value
of $\bar c_{LL}^\mu=-1.33$~\cite{DAmico:2017mtc}, and plot 
 the value of $g_L^{\mu\mu} - g_L^{sb}$ along this line on the ordinate. 
We see from the right-hand side plots in Fig.~\ref{fig:zprimesum} that the
na\"{i}ve model is not covered much at all by di-muon resonance searches at
the LHC, even at high luminosity, but that a 100 TeV 10 ab$^{-1}$ collider can cover all of the viable
parameter space where the $Z^\prime$ is narrow (we note that the sensitivity
at low masses is underestimated by our limit extrapolation technique,
as explained in Section~\ref{sec:extrapolationmethod}). However, the na\"{i}ve model
is a limiting case that underestimates both the potential sensitivity and the current
constraints for a more realistic model. We see in the left-hand plots that in a more complete $33\mu\mu$
model, a 14 TeV 1 ab$^{-1}$ LHC can cover a decent portion of the
viable parameter space and a 33
TeV LHC collider is sensitive to all of it.

\subsection{LQ sensitivity}

There are many dedicated experimental studies of LQs. For some recent
examples, CMS have searched for first and second generation LQs in
pair production~\cite{CMS-PAS-EXO-12-041,
  CMS-PAS-EXO-12-042,Khachatryan:2015vaa} and single
production~\cite{Khachatryan:2015qda} at 8 TeV centre of mass energy, while
ATLAS set limits on the pair production of third generation LQs using 7
TeV data~\cite{ATLAS:2013oea} and first and second generation LQs with
13 TeV~\cite{Aaboud:2016qeg}. A summary of LQ searches by ATLAS and
CMS can be found in Ref.~\cite{Romeo:2016abf}. LQs were recently reviewed in Refs.~\cite{Dorsner:2016wpm, Diaz:2017lit}. 

As the basis for our extrapolation, we take the 95 \% CL limits from the CMS 8
TeV search for a pair of second generation scalar LQs with 19.6 fb$^{-1}$ of
integrated luminosity~\cite{CMS-PAS-EXO-12-041}, focusing on the $\mu\mu jj$
channel in particular, as shown in Fig.~\ref{fig:diLQ}. The current limits
exclude masses up to $1070$ 
GeV, assuming a $100 \%$ branching fraction into a charged lepton and
quark. We note here that pair production proceeds through the strong
interaction and so limits coming from the experimental search may be phrased
as only depending on the LQ mass, 
once the assumption about its branching fraction is made.

\begin{figure}
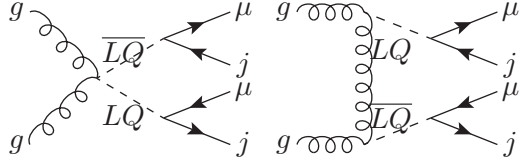

\begin{center}
\begin{axopicture}(180,50)(-5,0)
% 4 point diagram
\Gluon(0,0)(25,25){3}{3}
\Gluon(0,50)(25,25){3}{3}
\Line[dash](25,25)(50,40)
\Line[dash](25,25)(50,10)
\Line[arrow](50,10)(75,0)
\Line[arrow](75,20)(50,10)
\Line[arrow](50,40)(75,50)
\Line[arrow](75,30)(50,40)
\Text(-5,50)[c]{$g$}
\Text(-5,0)[c]{$g$}
\Text(80,50)[c]{$\mu$}
\Text(80,0)[c]{$j$}
\Text(80,20)[c]{$\mu$}
\Text(80,30)[c]{$j$}
\Text(35,35)[c]{$\overline{LQ}$}
\Text(35,10)[c]{$LQ$}

% 3 point diagram
\Gluon(100,0)(125,0){3}{3}
\Gluon(100,50)(125,50){3}{3}
\Gluon(125,50)(125,0){3}{6}
\Line[dash](125,50)(150,40)
\Line[dash](125,0)(150,10)
\Line[arrow](150,10)(175,0)
\Line[arrow](175,20)(150,10)
\Line[arrow](150,40)(175,50)
\Line[arrow](175,30)(150,40)
\Text(180,50)[c]{$\mu$}
\Text(180,0)[c]{$j$}
\Text(180,20)[c]{$\mu$}
\Text(180,30)[c]{$j$}
\Text(95,50)[c]{$g$}
\Text(95,0)[c]{$g$}
\Text(135,35)[c]{$LQ$}
\Text(135,10)[c]{$\overline{LQ}$}

\end{axopicture}
\end{center}
\caption{Example Feynman diagrams of LQ production at a hadron
  collider followed by
  subsequent 
  decay of each into $\mu j$. \label{fig:diLQ}}
\end{figure}

\begin{figure}
\centering
\includegraphics[width=0.49 \textwidth]{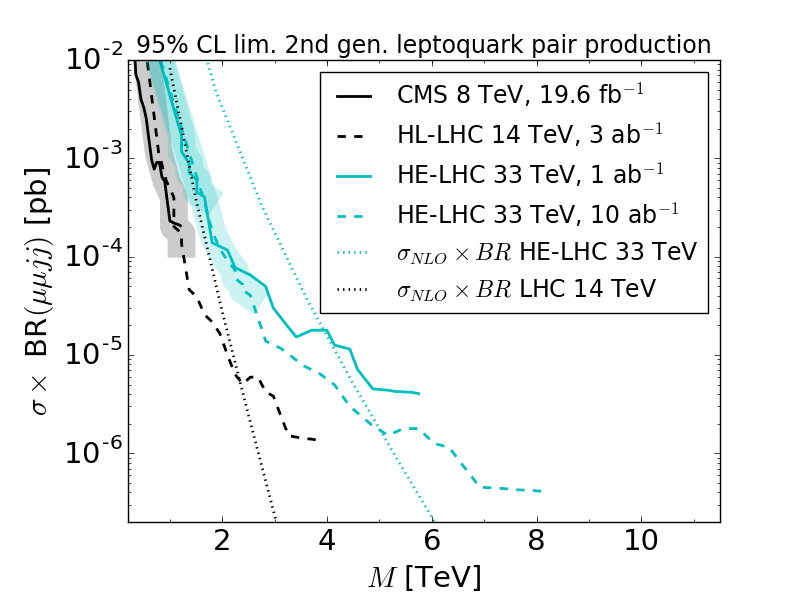}
\includegraphics[width=0.49 \textwidth]{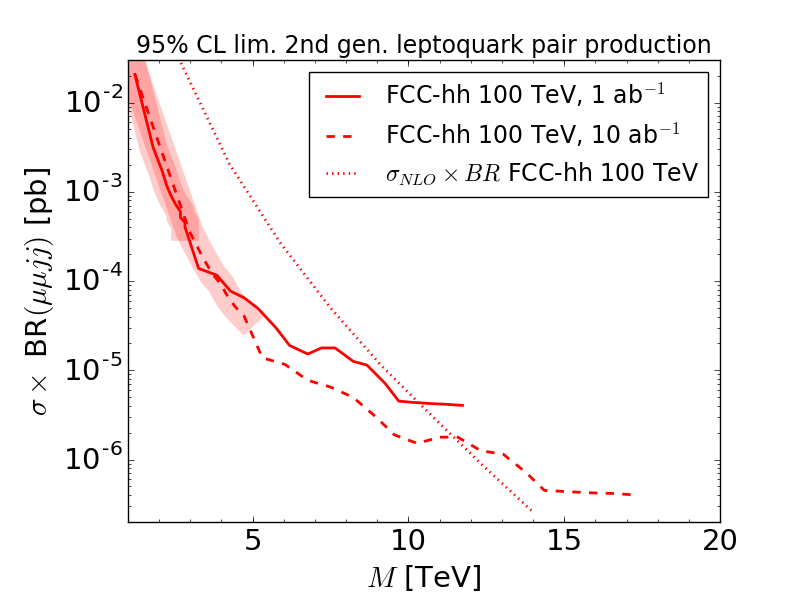}
\caption{Projected sensitivity of future colliders to di-LQ production
  (where each decays to a muon and a jet)
  for the luminosities and centre of mass energies
  given in the legend. We also show the scalar LQ cross-section times branching ratio
  {\em predicted}\/ for some future collider scenarios by the curves labelled
  $\sigma_{NLO} \times BR$. Shaded parts of the curve indicate the conservative extrapolation method at low masses that underestimates the actual limit.}
\label{fig:secondgenLQpairprodexclusion}
\end{figure}

Following the extrapolation procedure detailed in
Section~\ref{sec:extrapolationmethod}, we obtain the weighted sum of parton
pair luminosities for the dominant contributions to the background processes,
in this case $Z/\gamma^* + \text{jets}$ and $t\bar{t}$, then find the
equivalent mass at a future collider that gives the same number of background
events. The results for the projected limits are shown in
Fig.~\ref{fig:secondgenLQpairprodexclusion}. In the left hand plot, the
exclusion curve in solid black is the current CMS 8 TeV 
exclusion curve, while the dashed black line shows that the LHC reach can be
extended for 14 TeV at high luminosity (HL-LHC) with 3 ab$^{-1}$. The
cyan-coloured limits are for a potential high-energy upgrade to the LHC
(HE-LHC) that could reach up to 33 TeV centre of mass energy. The solid and
dashed lines represent 1 and 10 ab$^{-1}$ of integrated luminosities,
respectively. It appears that at low masses, the CMS 8 TeV analysis is more sensitive (when
phrased in terms of $\sigma \times BR$) than when the energy is upgraded to 14
TeV at the HL-LHC. This is an artefact of the arbitrariness in the starting
point of the extrapolated exclusion curve, as explained in Section~\ref{sec:extrapolationmethod}, where below this point lower luminosities can set limits at lower masses, though this conservative procedure underestimates the actual limit. The regions below the extrapolated starting point are shaded on top of their respective curves. On the right-hand side of
Fig.~\ref{fig:secondgenLQpairprodexclusion} we display the limits for a 100
TeV proton-proton future circular collider, the FCC-hh, at 1 (10) ab$^{-1}$
in solid (dashed) red.  

The dotted lines superimposed on both plots are theoretical calculations at
next-to-leading order for the LQ pair production process, using
the code of Ref.~\cite{Kramer:2004df}. Up to $\mathcal{O}(1)$ uncertainties,
we see that HL-LHC can exclude LQ masses up to $2$ TeV, while
HE-LHC can roughly double that to $4$ (5) TeV with 1 (10) ab$^{-1}$. At
FCC-hh the limits are improved by an order of magnitude with respect to
current searches, reaching exclusions up to $10$ and $12$ TeV for 1
and 10 ab$^{-1}$, respectively.

\begin{figure}
\centering
\includegraphics[width=0.55 \textwidth]{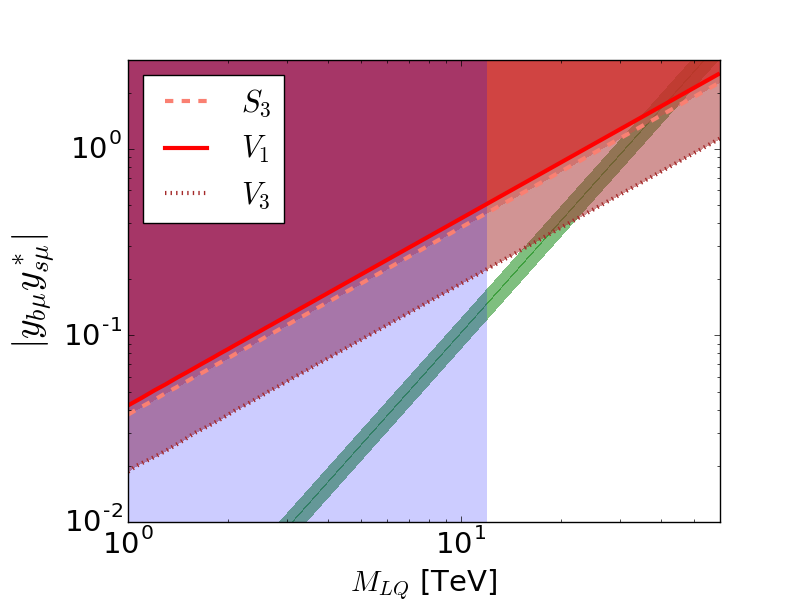}
\caption{Parameter space of the LQ on a log-log scale for couplings $|y_{b\mu} y^*_{s\mu}|$ vs mass in TeV. The green strip indicates the region compatible with the $B$-anomalies at $1 \sigma$. The different red-shaded regions are excluded by $B_s - \bar{B}_s$ mixing up to the solid red (dotted brown) line for the $V_1$ ($V_3$) vector LQ, and up to the dashed pink line for the $S_3$ scalar LQ, respectively. The region in blue up to $M_{LQ} \sim 12$ TeV is the projected $95 \%$ CL limit on scalar LQ pair production for FCC-hh at 100 TeV with 10 ab$^{-1}$.}
\label{fig:leptoquark}
\end{figure}

These projected bounds on the LQ mass are to be compared with the upper limit allowed by $B_s - \bar{B}_s$ mixing. The relevant four-fermion operator of the effective Lagrangian for this process can be written as
\begin{equation}
\mathcal{L}_{\bar{b}s\bar{b}s} = c_{LL}^{bb} \left(\bar{b}\gamma_\mu P_L s\right) \left(\bar{b}\gamma^{\mu} P_L s\right) + \text{h.c.}  \, .
\end{equation} 
The Wilson coefficient gets a contribution from the coupling combination $|y_{b\mu} y^*_{s\mu}|$ that is given by~\cite{Hiller:2017bzc}
\begin{equation}
c_{LL}^{bb} = k\frac{|y_{b\mu} y^*_{s\mu}|^2}{32\pi^2 M_{LQ}^2} \, ,
\end{equation}
where  $y=y_3,y_1,y_3'$ and $k = 5, 4, 20$ for the $S_3, V_1, V_3$ LQs,
respectively. Using this expression, together with Eq.~\ref{eq:cbarLLfromLQ} and the experimental
limit from $B_s - \bar{B}_s$ mixing that constrains $\bar{c}_{LL}^{bb}
\lesssim 1/(210 \text{ TeV})^2$~\cite{DAmico:2017mtc}, we obtain the parameter
space shown in Fig.~\ref{fig:leptoquark}. The couplings as a function of mass
are displayed on a log-log scale, and the green strip represents the parameter
space compatible with the $B$-anomalies at $1 \sigma$. The different shades of
red are excluded by $B_s - \bar{B}_s$ mixing for the $S_3, V_1, V_3$ LQs up to
the solid red, dotted brown, and dashed pink lines, respectively. We see that
the maximal values of the LQ masses allowed by $B_s - \bar{B}_s$ mixing
correspond to $M_{LQ} =37, 41, 18$ TeV for $S_3, V_1, V_3$, respectively. The
blue 
region shows the $95 \%$ CL limits for scalar LQs at a 100 TeV collider with
10 ab$^{-1}$, such as the FCC-hh. The pair production process for vector LQs
is more model-dependent (unlike scalar LQs, whose gluon interactions are fixed
by the $SU(3)_c$ gauge couplings) but is typically stronger than the scalar
case~\cite{Hewett:1993ks, Rizzo:1996ry, Hewett:1997ce}.

\begin{figure}
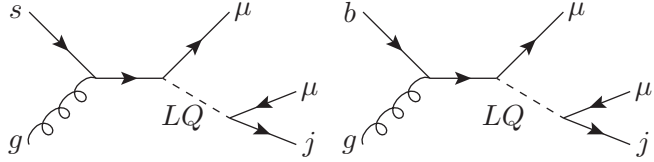

\begin{center}
\begin{axopicture}(230,50)(-5,0)
% Single production
\Gluon(0,0)(25,25){3}{3}
\Line[arrow](0,50)(25,25)
\Line[arrow](25,25)(50,25)
\Line[arrow](50,25)(75,50)
\Line[dash](50,25)(75,10)
\Line[arrow](75,10)(100,0)
\Line[arrow](100,20)(75,10)
\Text(-5,50)[c]{$s$}
\Text(-5,0)[c]{$g$}
\Text(105,20)[c]{$\mu$}
\Text(105,0)[c]{$j$}
\Text(80,50)[c]{$\mu$}
\Text(57.5,10)[c]{$LQ$}
% Single production
\Gluon(125,0)(150,25){3}{3}
\Line[arrow](125,50)(150,25)
\Line[arrow](150,25)(175,25)
\Line[arrow](175,25)(200,50)
\Line[dash](175,25)(200,10)
\Line[arrow](200,10)(225,0)
\Line[arrow](225,20)(200,10)
\Text(120,50)[c]{$b$}
\Text(120,0)[c]{$g$}
\Text(230,20)[c]{$\mu$}
\Text(230,0)[c]{$j$}
\Text(205,50)[c]{$\mu$}
\Text(177.5,10)[c]{$LQ$}
\end{axopicture}
\end{center}
\caption{Example Feynman diagrams of single LQ production at a hadron
  collider followed by its subsequent 
  decay into $\mu j$. Note that the LQ production cross-section depends upon
  its coupling to fermions, in contrast to the pair production cross-section
  depicted in Fig.~\protect\label{fig:diLQ}.\label{fig:singLQ}}
\end{figure}

The direct search sensitivity may also be extended to heavier LQs by
considering single LQ production~\cite{Belyaev:2005ew}, as in Fig.~\ref{fig:singLQ}. For
large enough couplings the limits may be be stronger than those obtained in
pair production~\cite{Khachatryan:2015qda}, but the product of the $b\mu$ and
$s\mu$ couplings must remain within the stringent bounds from $B_s -
\bar{B}_s$ mixing.

\begin{figure}
\centering
\includegraphics[width=0.49 \textwidth]{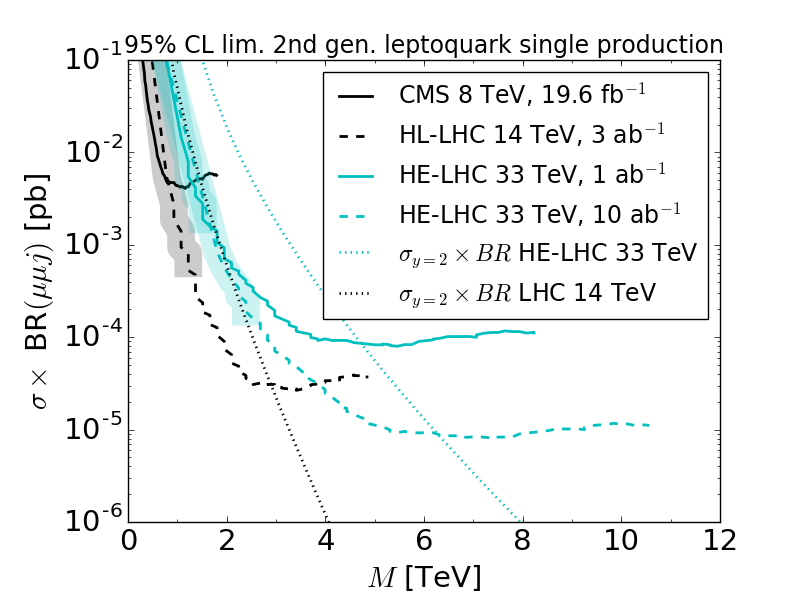}
\includegraphics[width=0.49 \textwidth]{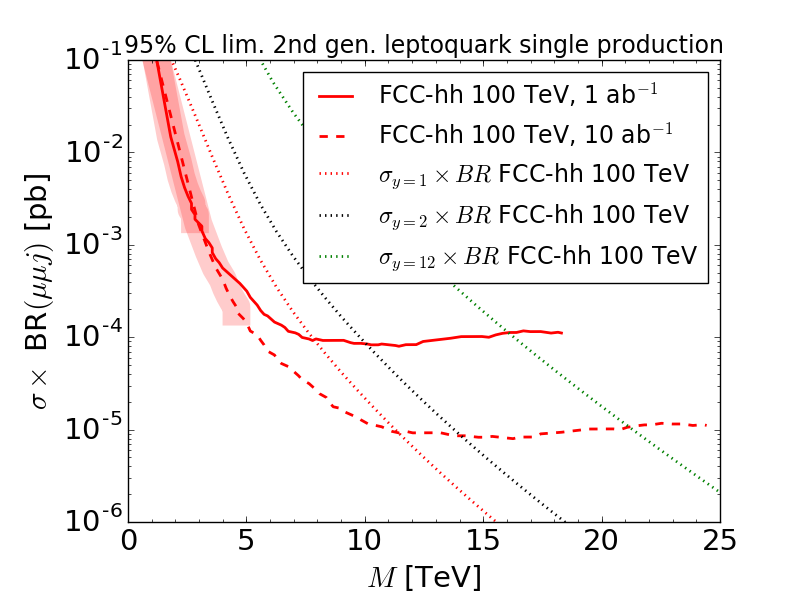}
\caption{Projected sensitivity of future colliders to single LQ production that decays to a muon and a jet,
  for the luminosities and centre of mass energies
  given in the legend. We also show the cross-section times branching ratio for some future collider scenarios by the curves labelled
  $\sigma_{y} \times BR$, where $y$ is the scalar LQ coupling to $b\mu$, set
  equal to the coupling to
  $s\mu$. Shaded parts of the curve indicate the conservative extrapolation method at low masses that underestimates the actual limit.}
\label{fig:secondgenLQsingleprodexclusion}
\end{figure}

We extrapolate the current limits from a direct search by CMS for a single
scalar LQ produced at 8 TeV with 19.6
fb$^{-1}$~\cite{Khachatryan:2015qda}. CMS places a bound of $M_{LQ} \lesssim
660$ GeV for a second generation LQ with coupling to $s\mu$ of order
unity. For our signal cross-section we also include a $b\mu$ coupling since we
take the $b$ quarks to be in the 5-flavour initial parton distribution
function 
NNPDF2.3LO ($\alpha_s(M_Z)=0.119$)~\cite{Ball:2012cx}. This re-sums the large logarithms of the initial state $b$-quark line. We integrate the partonic cross-section with the parton distribution functions using {\tt LHAPDF}~\cite{Buckley:2014ana}. The partonic cross-section at leading order for a scalar LQ $\phi$ is given by~\cite{Hewett:1987yg}
\begin{equation}
\hat{\sigma}(qg \to \phi l) = \frac{y^2 \alpha_S}{96 \hat{s}} \left( 1 + 6r - 7r^2 + 4r(r+1) \ln{r} \right) \, ,
\end{equation}
where $r = M_{LQ}^2 / \hat{s}$ and we set $y_{s\mu} = y_{b\mu} = y$ for
simplicity. This expression has been cross-checked with Fig. 8b of
Ref.~\cite{Khachatryan:2015qda} and found to agree within partonic
uncertainties. The extrapolated limits and production cross-sections for a
coupling and branching ratio set to 1 are displayed in
Fig.~\ref{fig:secondgenLQsingleprodexclusion}, with the same colour coding as
Fig.~\ref{fig:secondgenLQpairprodexclusion}. The signal cross-sections at 14
and 33 TeV are shown as dotted lines for $y=1$ in black and cyan respectively,
on the left plot. On the right we have the signal cross-section for 100 TeV
with $y=1, 2$, and $12$ in red, black, and green dotted lines respectively. We see that for couplings $y=1$, the limits are
comparable to pair production but slightly lower. On the other hand for $y=2$
the limits at 100 TeV go up to $ 15$ TeV for 10 ab$^{-1}$, extending to
$ 21$ TeV for $y \sim 4\pi$. The reach can be further extended for a model with additional quark couplings. Note however that in a realistic model the
product of $y_{b\mu}$ and $y_{s\mu}$ must still be anomaly-compatible within
the $B_s - \bar{B}_s$ mixing bounds shown in Fig.~\ref{fig:leptoquark}, so
that these limits only apply when one coupling is taken large with the other
small.

 \section{Conclusion}
 \label{sec:conclusion}
 
Some measurements of $B \to K^{(*)} l^+ l^-$ decays disagree with SM
predictions: using only theoretically clean quantities, the discrepancy 
on a Wilson coefficient with respect to the SM value is at around the
4$\sigma$ level~\cite{DAmico:2017mtc}. More specifically, the ratio of decays 
to muon pairs and electron pairs is predicted to be 1.0 in the SM, but is
measured to be lower than this value both for $K$ and $K^*$ in the final
state, each in two different bins of transferred 4-momentum
(squared). Moreover, the $4 \sigma$ pull from these clean observables on
a Wilson coefficient parameterising new physics is not only statistically
significant, but also in the same direction as another independent $4
\sigma$ pull due to other (less clean) observables. The combined significance
in a global fit is then significantly larger than 4$\sigma$.

Many authors have constructed bottom-up
models containing new particles in order to change the apparent predictions
and explain the discrepancies. In particular, it appears that lepton flavour
universality should be broken by the new particles, which should have chiral
interactions. At tree level, there are only two classes of new particle which
explain the discrepancies: flavourful $Z^\prime$s and LQs. We choose
these two cases to examine future hadron collider sensitivities: there are
other possibilities from particles which affect the decays at the loop level,
but because of the loop suppression, these particles should be a factor of
roughly $4
\pi$ lighter than the tree-level cases, and should therefore be easier to
detect directly by production in a collider. Studying the tree-level
possibilities is then conservative: if one shows that one can discover these,
it should be easier 
to find the lighter particles that are predicted by the loop effects.

We found that for the $Z^\prime$ models, a 100 TeV future collider will
essentially cover all of the 
parameter space that can explain the $B \to
K^{(*)} l^+ l^-$ decay data in a resonant di-muon search. 
Examining more complete models than the na\"{i}ve $Z^\prime$ model such as the
$33\mu\mu$ model,  we see that even a 33 TeV run of the LHC may cover all of the
relevant viable parameter space. The more complete models contain more model
dependence, but have stronger bounds and more coverage than the na\"{i}ve
model. 
One caveat to our analysis is that we only consider a narrow $Z^\prime$ with
width less than a tenth or so of the mass. This will not necessarily be the
case, but wide $Z^\prime$s invalidate the procedure we use to extrapolate
current bounds from scaling the background detailed in
Section~\ref{sec:extrapolationmethod}, and so 
require a more detailed simulation of backgrounds and signal at high
energies. Nevertheless, a wide $Z^\prime$ is theoretically disfavoured by the
large couplings required: they run into Landau poles. The $Z^\prime$ case is
summarised in Fig.~\ref{fig:zprimesum}, where 
expected 
sensitivities, bounds and the validity limit of our analysis are plotted for
various different future hadron collider assumptions. 

Coverage of the relevant LQ models is significant, but somewhat less
complete than the $Z^\prime$ models: whilst LQs of varying kinds up to
masses of $41$ TeV may explain the $B \to 
  K^{(*)} \mu^+ \mu^-$ decay data, we show that the expected sensitivity of
  di-LQ production into 
  a $\mu\mu j j$ final state reaches up to $ 12$ TeV for a scalar LQ. In
  more model-dependent cases the reach can be higher, as for example in pair
  production for vector LQs and single production for both vector and scalar
  LQs, which depend on a choice of couplings. We estimated the sensitivity of
  single scalar LQ production and found that $\mathcal{O}(1)$ couplings can
  reach a sensitivity up to LQ masses of around 21 TeV at the strong coupling
  limit. Whilst our extrapolation of current LHC limits is rather rough (one
  may expect an uncertainty of a factor of two in the cross-section times branching ratio for
  the limit due to PDF 
  uncertainties and different detector effects etc.), we estimate that this
  only results in an uncertainty on the sensitivity of around $\pm 1$ TeV when
  expressed in terms of the mass of leptoquarks or $Z^\prime$ particles (as
  evidenced by the steep model prediction curves in
  Figs.~\ref{fig:secondgenLQpairprodexclusion} and
  \ref{fig:secondgenLQsingleprodexclusion},  for example).

A potential loop-hole in our analysis would occur if one assumed the
existence of {\em multiple}\/ ($N$) mediators of the $c_{LL}^\mu$
operator. Under the assumption that each of the $N$ mediators ($Z^\prime$s or 
LQs) has an equivalent
mass and identical couplings, one obtains a contribution to $c_{LL}^\mu$ that
is 
proportional to either $N |y|^2 / M^2$ in the LQ case, or $N g_L^{sb}
g^{\mu\mu}_L/M_{Z^\prime}^2$ in the $Z^\prime$ case. For an identical effect
on the measured $b$ 
decays as in the unique mediator case, each of the $N$ mediators could
therefore be heavier by a factor $\sqrt{N}$ or more weakly coupled. 
The LHC mediator production cross-section falls with a power of the
mass that is significantly higher than two, resulting in weaker collider
sensitivity despite a factor of $N$ from the production of more new mediators.

Of course, there is always the possibility that the current discrepancy with SM
predictions is due to a fluke. In this case, our paper still serves a purpose,
estimating the reach of future colliders into particular flavourful
$Z^\prime$ or LQ models and demonstrating the interplay between indirect and direct searches for new physics.
In any case, 
new empirical data on the $B \to K^{(*)} l^+ l^-$ decays are expected from
Belle II and LHCb in the next few years. Ref.~\cite{Albrecht:2017odf} points
out that by 2020, 
the number of $b \bar b$ pairs produced inside the acceptance of LHCb 
should increase by a factor of 3.7 as
compared to those produced before and during 2012. For example, this would result in a
discovery of 
a non-SM effect beyond the 5$\sigma$ level in $R_K$, and close to
a 5$\sigma$ level effect in $R_{K^{*}}$ from LHCb data alone if the central
values were not to 
change from their current values~\cite{Albrecht:2017odf}. 

We have shown that there is significant coverage of all beyond the SM
explanations of the current anomalies in proposed future hadron
colliders. Thus, 
  if the signal significance of non-SM effects in $B \to 
  K^{(*)} \mu^+ \mu^-$ decays increases, so does this particular motivation
  for higher 
  energy future colliders\footnote{It does not follow that the motivation for going to higher energies then goes away if the anomalies vanish; there are many good reasons (that this margin is too small to contain) for furthering direct experimental exploration of the smallest scales for as long as we have the ability and curiosity to do so.}.

\section*{Acknowledgements}
We thank other members of the Cambridge SUSY Working Group and Michelangelo Mangano for
helpful advice
and comments. BCA thanks the Aspen Center for Physics for hospitality offered
while part of this work was carried out. TY thanks the Galileo Galilei Institute for hospitality and Nazila Mahmoudi for useful discussions. 
This work has been partially supported by STFC consolidated grants
ST/P000681/1 and ST/L000385/1 and by National Science Foundation grant PHY-1607611. TY is supported by a Junior Research Fellowship from 
Gonville and Caius College, Cambridge.

\bibliographystyle{JHEP-2}
\bibliography{futureAnom}
 
\end{document}